\documentclass[structabstract]{./aa}
\usepackage{graphicx,epsfig,natbib}
\usepackage{txfonts}
\usepackage{setspace}
\usepackage{threeparttable}

\usepackage{natbib}
\bibpunct{(}{)}{;}{a}{}{,} 

\begin{document}

\title{The quiet Sun average Doppler shift of coronal lines up to 2~MK}

\author{Neda Dadashi\inst{1,2}
  \and Luca Teriaca\inst{1}
  \and Sami K. Solanki\inst{1,3}}

\offprints{N. Dadashi, \email{dadashi@mps.mpg.de}}

\institute{Max-Planck-Institut f\"{u}r Sonnensystemforschung, 37191 Katlenburg-Lindau, Germany
\and Institut f\"{u}r Geophysik und extraterrestrische Physik, Technische
Universit\"{a}t Braunschweig, Mendelssohnstr. 3, D-38106 Braunschweig, Germany
\and School of Space Research Kyung Hee University, Yongin, Gyeonggi-Do,
446-701, Korea}

\date{Received , ; accepted , }

\abstract
{The average Doppler shift shown by spectral lines formed from the
chromosphere to the corona reveals important information on the mass and energy
balance of the solar atmosphere, providing an important observational constraint
to any models of the solar corona.
Previous spectroscopic observations of vacuum ultra-violet (VUV)
lines have revealed a persistent average wavelength shift of lines formed at
temperatures up to 1~MK. At higher temperatures, the behaviour is still
essentially unknown.}
{Here we analyse combined SUMER (Solar Ultraviolet Measurements of
Emitted Radiation)/SoHO (Solar and Heliospheric Observatory) and
EIS (EUV Imaging Spectrometer)/Hinode observations of
the quiet Sun around disk centre to determine, for the first time, the average
Doppler shift of several spectral lines formed between 1 and 2~MK, where the largest part
of the quiet coronal emission is formed.}
{The measurements are based on a novel technique applied to EIS spectra to
measure the difference in Doppler shift between lines formed at different
temperatures. Simultaneous wavelength-calibrated
SUMER spectra allow establishing the absolute value at the reference
temperature of $T\approx$ 1~MK.}
{ The average line shifts at 1~MK$~<T<1.8$~MK
are modestly, but clearly bluer than those observed at 1~MK.
By accepting an average blue shift of
about $(-1.8\pm0.6)~\rm km~s^{-1}$ at 1~MK (as provided by SUMER measurements),
this translates into a maximum Doppler shift of
$(-4.4\pm2.2)~\rm km~s^{-1}$ around 1.8~MK.
The measured value appears to decrease to about $(-1.3\pm2.6)~\rm km~s^{-1}$
at the Fe~{\sc xv} formation temperature of 2.1~MK.}
{ The measured average Doppler shift between 0.01 and 2.1~MK, for which we
provide a parametrisation,
appears to be qualitatively and roughly quantitatively
consistent with what foreseen by 3-D coronal models where heating is
produced by dissipation of currents induced by photospheric motions and by
reconnection with emerging magnetic flux.}


\authorrunning{Dadashi, Teriaca, Solanki}

\titlerunning{The quiet Sun average Doppler shift of coronal lines}

\maketitle

\section{Introduction}
Ultraviolet observations of the Sun near disk centre show a systematic
net red shift of emission lines from the transition region (TR).
Several vacuum ultra-violet (VUV: 16 to 200~nm)
spectrometers with different spatial resolutions such as HRTS and those
flown on Skylab, OSO~8, and SMM have revealed this phenomenon since the 1970s
\citep[e.g.,][]{Doschek76, Hassler91, Brekke93, Achour-etal:95}.
Systematic red shifts have also been observed in spectra of late type
stars \citep[e.g.,][]{Ayres88,Wood-etal:96, Pagano-etal:04},
indicating that this phenomenon is not unique to the Sun.

In these earlier investigations (before the launch of SoHO in 1995),
the magnitude of the red shift was found to increase with temperature,
reaching a maximum at $T=0.1$~MK and then decreasing toward higher
temperatures.
However, these observations were, with few exceptions, restricted to
temperatures below 0.2~MK.

Observations from SOHO extended the observable temperature range up to about
1~MK. In the quiet Sun near disk centre, Doppler shifts in the range of
10 to 16~km~s$^{-1}$ were confirmed in lines formed from $T=$~0.1~MK to
0.25~MK \citep[e.g.,][]{Brekke97, Chae98, PetJud99, Teriaca99a}.

At higher temperatures (0.6 to 1~MK), the interpretation of the
measurements proved to be more controversial.
\citet{Brekke97} and \citet{Chae98} reported red shifted (downward) average
Doppler velocities of 6.0 and 3.8~km~s$^{-1}$ for
Mg~{\sc x}~$\lambda$625 ($T=$~1.0~MK) and of 5.0 and
5.3~km~s$^{-1}$ for Ne~{\sc viii}~$\lambda$770 ($T=$~0.63~MK).
These Doppler velocities were derived by using the
reference rest wavelengths of 770.409~\AA\ and 624.950~\AA\
compiled by \citet{Kelly87} \citep[based on solar observations,
see discussion in][]{PetJud99}, affected by high
uncertainties (high in relation to the Doppler shifts under discussion).

\citet{PetJud99} and \citet{Dam99a, Dam99b} took another approach and obtained
accurate rest wavelengths for the above lines by assuming that the average
Doppler velocity is around zero above the solar limb (motions along the line of
sight cancel out on average in an optically thin plasma). With these new rest
wavelengths, average {\it blue shifts} of a few km~s$^{-1}$ were found for both
lines near disk centre \citep{Peter99, PetJud99, Dam99a, Dam99b, Teriaca99a}.
These results suggest that there is a transition from red to blue shifts
above 0.5~MK. This transition is an important observational constraint to
distinguish between different models of the solar transition region.

Several hypotheses to explain the observed shifts of TR and coronal
lines have been discussed in the literature.
Since spicules carry an upward mass flux about 100 times larger than carried
out by the global solar wind, it was suggested that the TR emission is dominated by
downflowing, cooling plasma injected into the corona by spicules
\citep{Pneumankopp78, AthayHolzer:82, Athay84}.
Lack of direct observational evidence \citep{Withbroe:83}
and failure of theoretical models to reproduce observations
\citep{Mariska87} have led to a general dismissal of this hypothesis.
However, recent analysis of spectroscopic and imaging VUV data with high
sensitivity and spatial resolution has revived interest in the role
of spicules in the mass and energy balance of the solar corona
\citep[e.g.,][]{DePontieu-etal:2009, McIntosh-etal:2009b, McIntosh-etal:2009a,
DePontieu-etal:2011}.

 Much effort has also been put into modelling the effect of energy deposition
(e.g., mimicking the effect of magnetic reconnection in
different regions of the solar atmosphere) on line profiles.
\citet{Cheng94} showed that impulsive energy release in the photosphere
generates a shock train that propagates upward and lifts and heats the chromosphere and
TR, giving rise to an average downward material velocity.
However, \citet{Hasteen94} demonstrated that, contrary to observations, the
resulting line emission is nevertheless blue shifted.
Instead of upward propagating waves, \citet{Hasteen93} interpreted the observed
red shift as caused by downward propagating acoustic waves. Nanoflares
releasing relatively small amounts of energy near the top of the magnetic
loops give rise to disturbances running downward along the loop legs.
In a later study \citet{Hansteen97} included the effect
of reflection in the chromosphere of downward propagating MHD waves.
This increases the red shifts in the TR, although coronal blue shifts
become too high.
An investigation by \citet{Jud98} lends observational
support to the picture of downward running waves or disturbances.

More generally, models of energy deposition at TR temperatures in 1-D loops
have managed to reproduce the observed TR red shifts and low corona
blue shifts relatively well \citep{Teriaca99b, Spadaro-etal:06}.

Recently, three-dimensional (3-D) comprehensive numerical models
of the solar atmosphere (from photosphere to corona) have been developed.
In the model described by \citet{Peter2004, Peter2006}, coronal heating is
caused by the Joule dissipation of the currents that
are produced when magnetic field is stressed and braided by photospheric
motions. This model does produce red shifts at TR temperatures but also
in the low corona, where blue shifts are instead observed.
\citet{Hansteen2010} extended that work and added the effects of episodic
injections of emerging magnetic flux that reconnects with the existing field.
Rapid, episodic heating of the upper chromospheric plasma
to coronal temperatures naturally produces downflows in TR lines
and slight upflows in low coronal lines.
As suggested by the authors, the localised heating events occurring
in the chromosphere generate high-speed flows in a wide range of
temperatures that may be related to the high-speed upflows that have been
deduced from significant blueward asymmetries in EIS/Hinode observations
of many TR and coronal lines \citep{Hara-etal:2008, DePontieu-etal:2009,
McIntosh-etal:2009b, McIntosh-etal:2009a} that have been linked to spicules
\citep[i.e.,][]{DePontieu-etal:2011}.

With the launch of the Extreme Ultraviolet Imaging Spectrograph
\citep[EIS,][]{Culhane2007, Korendyke2006} aboard Hinode,
high spectral and spatial resolution observations in intense spectral lines
formed at temperatures above 1~MK have finally become available.
In this work, by using combined SUMER (Solar Ultraviolet Measurement of
Emitted Radiation)/SoHO and EIS/Hinode observations of the quiet Sun
and a novel technique, we have characterised the
Doppler shift versus temperature behaviour up to 2.1~MK.

In Section 2, we describe the SUMER data and their analysis. In addition, we
determine the Doppler velocity for
N~{\sc v}~$\lambda$1238, N~{\sc v}~$\lambda$1242, and
Mg~{\sc x}~$\lambda$625
lines in the quiet Sun near disk centre.
We also discuss the EIS/Hinode data, their analysis, and their
co-alignment with the SUMER data.
In Section 3 we present our method to measure the average Doppler velocity of
hot coronal ions using SUMER/SoHO and EIS/Hinode data. Section 4 and 5 present
results and conclusions, respectively.

\section{Observations and data analysis}
\begin{figure}[t!]
   \centering
   \includegraphics[trim=20mm 0mm 20mm 0mm, clip, width=70mm]{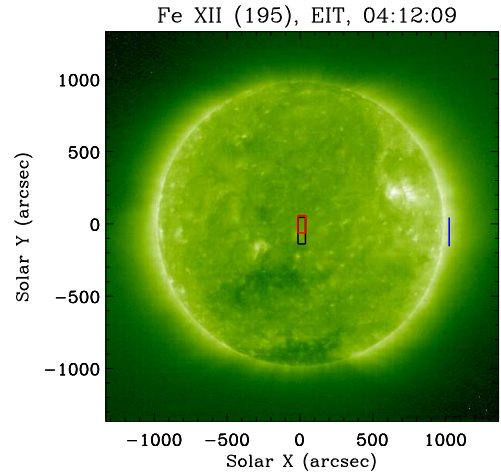}
    \caption{Fe~{\sc xii} 195~\AA\ EIT image. The areas observed on the disk
    by SUMER (smaller red box) and EIS (larger black box) are shown.
    They are located near disk centre in a quiet region.
    The off-limb pointing of the EIS slit (about 50~$\arcsec$ above the west limb)
    is indicated by the vertical black line. }
   \label{fig:context}
 \end{figure}

Spectra of a quiet region near disk centre were taken on 6 April 2007
from 00:00 to 06:00 UTC during the first EIS-SUMER campaign (HOP 3).
Figure~\ref{fig:context} shows the Sun on the day of observations as
recorded by the Extreme Ultraviolet Telescope
\citep[EIT,][]{Delaboudin-etal:95} aboard SOHO. The quiet Sun areas observed
by SUMER and EIS are visible near disk centre.
%
\subsection{SUMER data: Doppler shift of Mg~{\sc x} and N~{\sc v}}
SUMER is a normal incidence spectrograph operating over the wavelength
range 450~\AA\ to 1610~\AA.
It is a powerful UV instrument capable of making
measurements of bulk motions in the chromosphere, TR, and low
corona with a spatial resolution of 1$\arcsec$ across and 2$\arcsec$ along the
slit and a spectral scale of $\approx$~43~m\AA/pix at 1240~\AA\ (first order)
\citep{Wilhelm95, Lemaire97}, which is high enough to measure line shifts down
to 1~km~s$^{-1}$. Spectral images have been decompressed, wavelength-reversed,
corrected for dead time, flat-field, and detector electronic distortion.

Since SUMER does not have an on-board calibration source, we derive the
wavelength scale by using a set of chromospheric lines assumed to have
negligible or low average Doppler shift on the quiet Sun.
In short, calibrating takes three main steps:

\begin{enumerate}
  \item Identify the spectral lines by using the preliminary wavelength scale.
  \item Find the exact pixel position of the lines by fitting Gaussian curves.
  \item Perform a polynomial fit to find the dispersion relation.
\end{enumerate}

To perform an accurate calibration we have to choose fairly strong and unblended
lines from neutral and singly ionised atoms. For instance, Si~{\sc i},
S~{\sc i}, and C~{\sc i} transitions, occurring at temperatures between 6500
and 10000~K, are very suitable calibration lines, since lines formed at such
temperatures show, on average, no substantial Doppler shifts
\citep{Samain:1991}.

The SUMER data analysed here consist of a raster formed by 40 slit positions
with a step size of $1.87\arcsec$ taken between 03:26 and 04:31 UTC by exposing
through the $1\arcsec \times 120\arcsec$ slit for 100~s.
Recorded spectra cover the 1210~\AA\ to
1270~\AA\ range containing the N~{\sc v} lines at 1238.8~\AA\ and
1242.8~\AA\ (log($T$/[K])=5.24) and the Mg~{\sc x} line at 625~\AA\
(log($T$/[K])=6.00). The above spectral range also contains a forbidden line
of Fe~{\sc xii} at 1242~\AA\ that would be ideal for a combined study with
EIS. However, on the quiet Sun this line is weak and blended with
chromospheric lines from Ca~{\sc ii} and S~{\sc i} \citep[see][]{Dam99a}
and, for this reason, it is not possible to derive its average Doppler shift with
the accuracy needed for our study.

\begin{figure}
  \hspace{0.1cm}
  \resizebox{8.5cm}{!}{\includegraphics{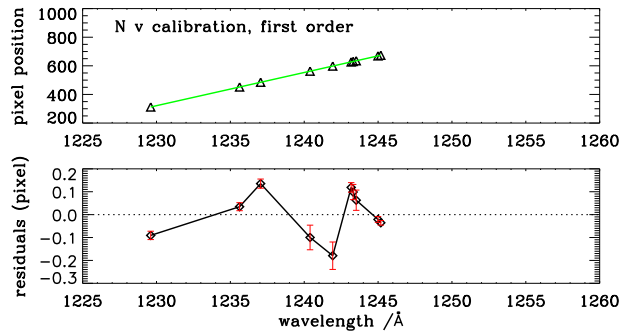}}
  \vspace{-1.5cm}
  \caption{Upper panel: first-order polynomial fit to pixel positions versus rest
               wavelengths of 10 suitable reference lines around the
               N~{\sc v} lines. Lower panel: fit residuals are under 0.2~pixels
               (less than 2~km~s$^{-1}$).}
  \label{pic}
 \end{figure}

After checking that there are no significant instrumental drifts in
the position of the spectral lines (either along the slit either in time
across the raster) we have obtained a high signal to noise spectrum to be
calibrated in wavelength.
We have selected ten suitable reference lines in the part of the spectrum which
contains the N~{\sc v} lines. A first-order polynomial fit of their
positions in pixels versus their laboratory wavelengths, leaves residuals of
less than $\pm$ 0.2 pixels (less than 2~km~s$^{-1}$, see Fig.~\ref{pic}).

\begin{figure}[htp]
   \centering
   \includegraphics[width= 0.41 \paperwidth]{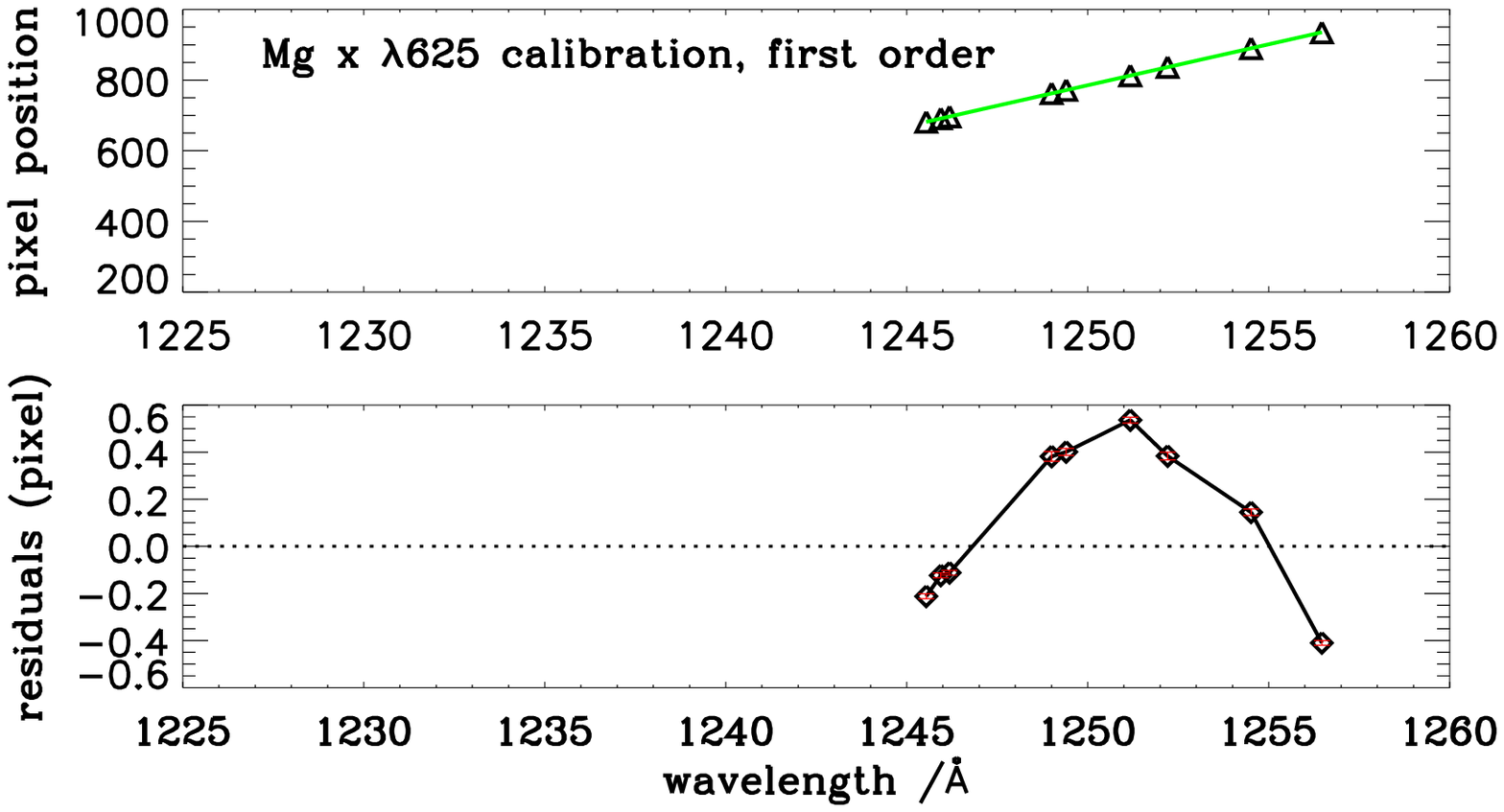}
   \includegraphics[width= 0.41 \paperwidth]{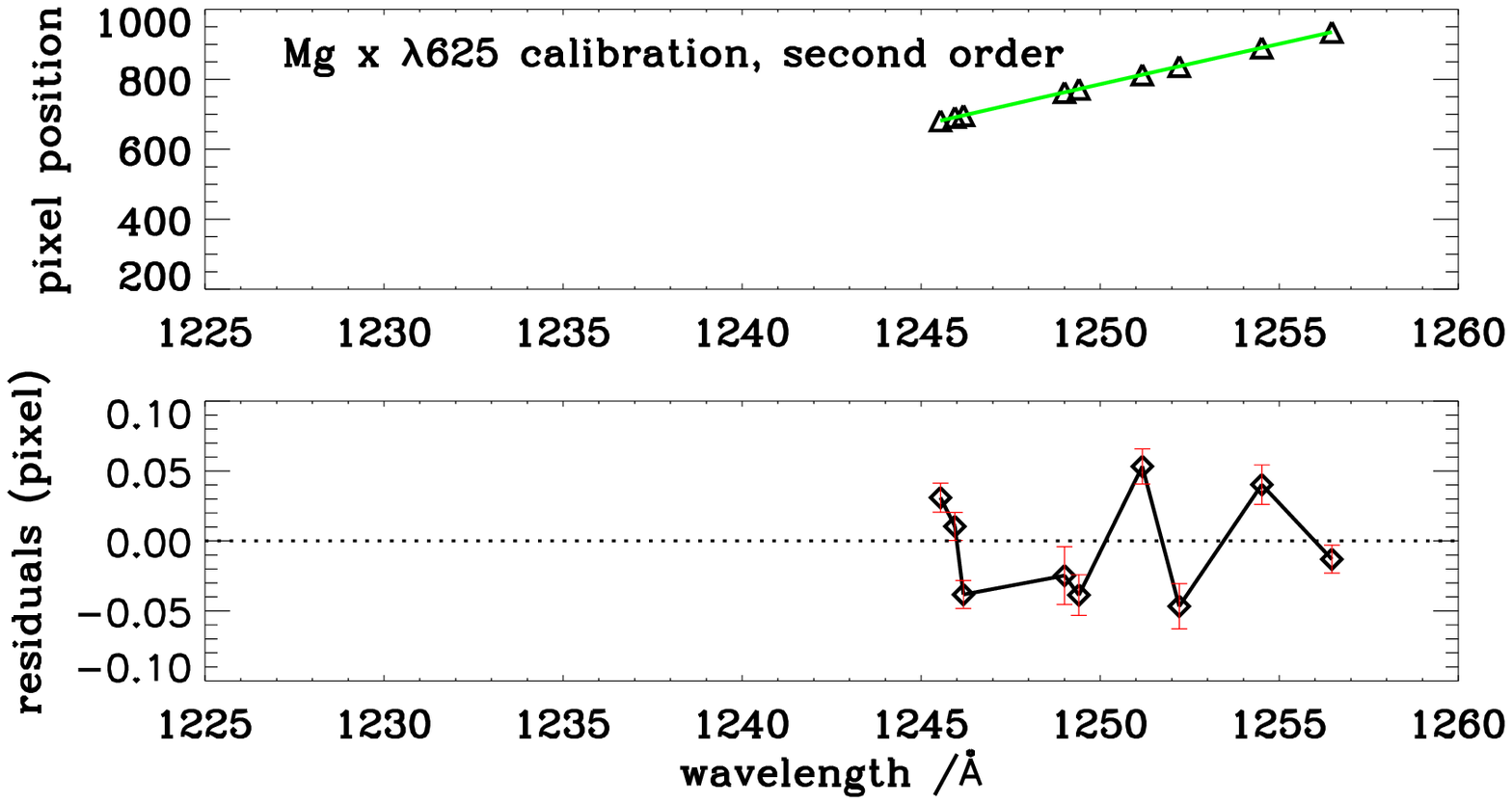}
    \vspace{-1.cm}
    \caption{ Top pair of panels: first-order polynomial fit to pixel positions versus
              rest wavelengths of 9 suitable reference lines around the Mg~{\sc x}
              lines. Residuals are large, up to 0.6~pixels.
              Bottom pair of panels: second-order polynomial fit to pixel positions
            versus rest wavelengths. Residuals lie within $\pm$ 0.05 pixels.}
   \label{pic0}
 \end{figure}

For the part of the spectrum containing the Mg~{\sc x} line, we have
selected nine suitable reference lines. If we use a first-order polynomial fit
to the line positions, the residuals up to $\pm$ 0.6 pixels (top panel
of Fig.~\ref{pic0}).
The residuals clearly show the need for a second-order polynomial fit
that yields residuals of $\pm$ 0.05 pixels (bottom panel of Fig.~\ref{pic0}).
The need for the higher order fit comes most likely from residuals in the
correction of the electronic distortion of the detector image.

From the above analysis, for the N~{\sc v}~$\lambda$1238 and $\lambda$1242
lines we find net red shifts of (9.9 $\pm$ 0.6)~km~s$^{-1}$ and
(12.1 $\pm$ 0.6)~km~s$^{-1}$ by assuming the rest wavelengths
to be 1238.800~\AA\ and 1242.778~\AA\ \citep{Edl34}.
These velocities agree with those from lines formed at similar temperatures
reported in the literature \citep[e.g.,][]{Brekke97,Chae98,PetJud99, Teriaca99a}.

In the case of the Mg~{\sc x}~$\lambda$625 line (observed in the second-order
of diffraction), we use 624.967~\AA~ as rest wavelength
(average between the values of (624.965 $\pm$ 0.003)~\AA\ \citep{Dam99a} and
(624.968 $\pm$ 0.007)~\AA\ \citep{PetJud99}),
obtaining a Doppler velocity of ($- 1.8 \pm 0.6$)~km~s$^{-1}$
(Table~\ref{table:1}). With the above rest wavelength, the
Doppler shifts measured in Mg~{\sc x} by \citet{Brekke97} and \citet{Chae98}
would result in velocities of about 0~km~s$^-1$ and $-2$~km~s$^-1$,
respectively.

\citet{Brekke97} discuss a possible blend in the red wing of
the Mg~{\sc x}~$\lambda$625 line with a first-order Si~{\sc ii} 1250.089~\AA\
line and conclud that it has a negligible effect on the derived line shifts
in spectra taken on the KBr-coated part of the SUMER detector.
\citet{Teriaca-etal:2002} also discuss the relevance of blending from first-order
lines to the Mg~{\sc x} line in the quiet Sun, concluding that they account
for about 10\% to 15\% of the signal in the observed profile when observed on KBr
(mostly from the Si~{\sc ii} 1250.089~\AA\ line) and to 1\% to 2\% when observed
on the uncoated detector (due to the low sensitivity to first-order lines with
respect to the KBr-coated parts).
Since our observations were obtained on the uncoated part of the SUMER
detector, we consider blends from first-order lines as entirely
negligible.

Blends from second-order lines such as S~{\sc x} 624.70~\AA,
O~{\sc iv} 624.617~\AA, and 625.130~\AA\ are potentially more dangerous.
However, these lines are far enough from the centre of the Mg~{\sc x}
line and would produce detectable ``shoulders'' in the observed profile.
Moreover, constrained double
Gaussian fits show no significant improvement in line fitting over
fits with a single Gaussian. This speaks against significant
blending in the Mg~{\sc x}~$\lambda$625 line also from second-order lines.

\begin{table}
\caption{Doppler velocity of N~{\sc v} and Mg~{\sc x} lines:}
\label{table:1}
\centering
 \begin{tabular}{c c c c}
  \hline \hline & \mbox{}\\[-1.5ex]
ions                & rest wavelength &     $v$              & $\delta v$         \\
(log ($T$/[K]))  & [\AA]                 &[km~s$^{-1}$]  & [km~s$^{-1}$] \\
  \hline & \mbox{}\\[-1.5ex]
   N~{\sc v} (5.24) & 1238.800 & 9.9  & 0.6 \\
   N~{\sc v} (5.24) & 1242.778 & 12.1 & 0.6 \\
   Mg~{\sc x} (6.00) & 624.967 & -1.8 & 0.6 \\
   \hline
 \end{tabular}
 \end{table}
\subsection{EIS data: analysis and co-alignment}
The EIS instrument on Hinode produces high-resolution stigmatic spectra in
the wavelength ranges of 170 to 210~\AA\ and 250 to 290~\AA.
The instrument has $1^{\arcsec}$ spatial  pixels and 0.0223~\AA\ spectral
pixels.
More details are given by \citet{Culhane2007} and \citet{Korendyke2006}.

Our EIS data consist of four raster scans and one sit-and-stare sequence.
The four scans were taken at disk centre on the same target area as observed by
SUMER, whereas the sit-and-stare sequence was obtained above
the limb. EIS rasters are obtaining by scanning from
west to east, in the opposite direction to the SUMER data discussed here.
Raster 4 is the closest in time to the SUMER scan.
Data were reduced by using the software provided within SolarSoft.

\begin{figure}
  \resizebox{8.5cm}{!}{\includegraphics{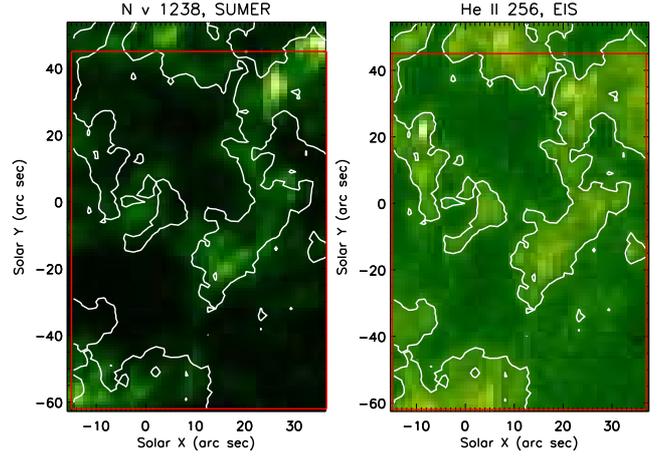}}
  \caption{Radiance maps of the common area of study in
            N~{\sc v}~$\lambda$1238~\AA\ (SUMER, left) and
          He~{\sc ii}~$\lambda$256~\AA\ (EIS, right).
          Contours of He~{\sc ii}~$\lambda$256 are plotted on
          both maps. The red boxes show the area in common to SUMER
          and the short and long wavelength ranges of EIS.}
  \label{pic1}
 \end{figure}

After accounting for the offset between the short and the long wavelength
ranges of EIS,
we use pairs of radiance maps obtained in lines from similar (at least not
too different) temperatures to perform a co-alignment between scans made by
the two instruments with an accuracy of roughly $1^{\arcsec}$
(see Fig.~\ref{pic1}). The employed line pairs include
N~{\sc v}~$\lambda$1238 (log($T$/[K])=5.24, SUMER) and He~{\sc ii}~$\lambda$256
(log($T$/[K])=4.7, EIS), as well as Mg~{\sc x}~$\lambda$625 (log($T$/[K])=6.00,
SUMER) and Fe~{\sc x}~$\lambda$184 (log($T$/[K])=6.00, EIS).

Since the wavelength range covered by EIS does not contain any suitable
neutral or singly ionised line (He~{\sc ii}~$\lambda$256, besides being
optically thick, is formed at too high a temperature to be un-shifted),
an absolute wavelength calibration is not possible using the EIS data alone.
We, therefore, take the approach of finding spectral lines recorded by the two
instruments that are formed roughly at the same temperature by assuming that
their average quiet Sun velocities are the same. Specifically, we assume
that the average velocities of Mg~{\sc x} and Fe~{\sc x} ions are the same
in the common observed area and equal to ($-1.8 \pm 0.6$)~km~s$^{-1}$.

The above assumption requires some further consideration as it cannot
be immediately excluded that the different ions have different
velocities. In fact, mechanisms that can selectively heat and
accelerate ions, such as ion-cyclotron resonance absorption of Alfv\'{e}n
waves \citep[see, e.g.,][]{1997SoPh..171..363T,1997A&A...319L..17M},
have long been considered as candidates for the solar wind acceleration.
This mechanism is usually considered for low-density, open field regions
such as coronal holes. However,
\citet{2003A&A...411L.481P} claim that it may also
operate at the base of open funnels in the quiet Sun, some 2 to 10~Mm
above the photosphere. Since Fe~{\sc x} and Mg~{\sc x} have substantially
different mass-to-charge ratios (6.2 and 2.7, respectively),
the Fe~{\sc x} ions could be accelerated more (due to the lower
gyro-frequency) and at lower heights
(due to the expected fall-off of the magnetic field strength with height)
than the Mg~{\sc x} ions.
Thus, we may be underestimating the Fe~{\sc x} average outflow speed
by taking it to be equal to that of Mg~{\sc x}.
On the other hand, at slightly higher temperatures, we obtain very
consistent results from lines of Fe~{\sc xii} and Si~{\sc x} that
also have different mass-to-charge ratios (5.1 and 3.1), although
the difference is smaller.

It should also be noted that Fe and Mg have very similar first ionisation
potentials (FIP), excluding eventual differences in the average flows due
to the fractionation processes believed to occur in the chromosphere
\citep[see, e.g.,][]{2004ApJ...614.1063L}.

From the above considerations, we conclude that assuming the average velocity
of Fe~{\sc x} ions to be equal to that of Mg~{\sc x} ions is justified and
should not result in a systematic error that would be large enough to change
our results and conclusions substantially.
Therefore, in the following, we take Fe~{\sc x}~$\lambda$184 as our
reference line with respect to which the average Doppler shift of the
remaining spectral lines recorded by EIS is determined.

%
\section{A method of measuring the Doppler velocity of EIS coronal lines}
\label{sec:method}
\begin{figure}
  \resizebox{9.cm}{!}{\includegraphics{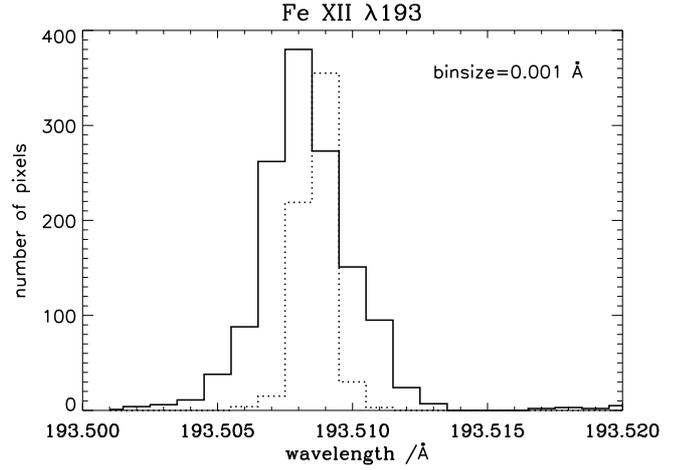}}
  \caption{The histograms of line position in wavelength for
               Fe~{\sc xii}~$\lambda$193 line off-limb (dotted line) and
               at disk centre (solid line). The off-limb histogram is narrower
               than the on-disk one because {\bf net mass} motions are
             essentially absent.}
 \label{pic3_}
 \end{figure}

\begin{figure}
  \resizebox{9.cm}{!}{\includegraphics{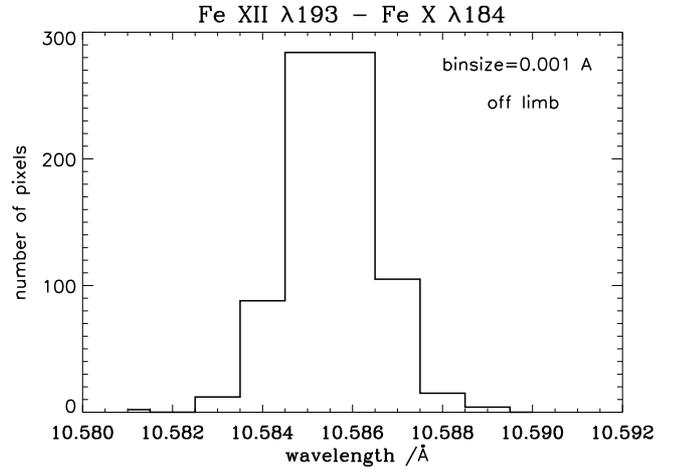}}
  \caption{The histogram of distance in wavelength between
               Fe~{\sc xii}~$\lambda$195 line and
               the reference line (Fe~{\sc x}~$\lambda$184) off-limb.
               The histogram is nearly Gaussian and with a sigma of
               $\approx$ 1.3~km~s$^{-1}$, that can be taken as an estimate
             of the instrument accuracy in measuring the distance in wavelength.}
  \label{pic4}
 \end{figure}

Above the limb we assume that spectral lines have, on average, no Doppler shift
(motions out of the plane of sky cancel out on average).
This is verified by plotting histograms of velocity (with respect to the average)
and histograms of the distance in wavelength between a line and the reference
Fe~{\sc x}~$\lambda$184 line.
These histograms are very narrow, nearly Gaussian, and have an FWHM of
3~km~s$^{-1}$ to 4~km~s$^{-1}$ (Figs. \ref{pic3_}, \ref{pic4}).

From the above assumptions\footnote{To be precise, the quantity
$\overline {\Delta \lambda _{off}}$ we measure from the
histogram of the distance in (off-limb) wavelength should be corrected for the effect
of solar rotation before being used at disk centre. This requires the right-hand side of
Eq.~\ref{eq:doff} to be multiplied by ${1+v_{\rm rot}/c}$. In the case of the Sun,
$v_{\rm rot}\approx 2$~km~s$^{-1}$ and the term is practically equal to one.}
\footnote{Also the gravitational red shift plays no role here. In fact,
with EIS we are comparing distances between spectral lines on the disk and above the
limb, and the effect, because the gravitational red shift is the same at the two places, cancels out.
In the case of the SUMER data, the Doppler shift are measured with respect to the chromosphere,
which is assumed to be at rest.}, the average of the off-limb line-distance histogram is
then

\begin{equation}
 \overline {\Delta \lambda _{off} }  = \,\lambda _0  - \lambda _0^{Ref},
  \label{eq:doff}
 \end{equation}

\noindent where $\lambda_0^{Ref}$ and $\lambda_0$ are the rest wavelengths of the
reference line and of the line whose shift we want to measure.

On the disk we measure the quantity
$\Delta \lambda  = \,\lambda  - \lambda^{Ref}$
at each pixel or group of pixels (where $\lambda^{Ref}$ is the wavelength
of the reference line).
This can be written as
\begin{equation}
\Delta \lambda  = \,\lambda _0 + \delta \lambda \, - \,\lambda _0^{Ref} \, - \,\delta \lambda^{Ref}  = \,\overline {\Delta \lambda_{off}} \, + \delta \lambda \, - \,\delta \lambda^{Ref},
\label{eq:dl1}
\end{equation}

\noindent where $\delta \lambda$ and $\delta \lambda^{Ref}$ are the line Doppler shifts.
Considering that $v = c\frac{{\delta \lambda }}{{\lambda _0}}$, we have
\begin{equation}
\Delta \lambda  = \,\overline {\Delta \lambda _{off} } \, + \frac{v}{c}\,\lambda _0 \, - \frac{{v^{Ref}}}{c}\,\lambda _0^{Ref}.
\label{eq:dl2}
\end{equation}

\noindent If we call $\delta v = v - v^{Ref}$, the difference between the
velocities measured in the two lines, then $\delta v$ becomes
\begin{equation}
\delta v\, = \,\,\frac{c}{{\lambda _0 }}\left( {\Delta \lambda \, - \,\overline {\Delta \lambda _{off} } \left( {1 + \frac{{v^{Ref} }}{c}} \right)} \right).
\label{eq:dv}
\end{equation}

\noindent Considering now the average value of $v^{Ref}$, $\overline {v^{Ref} }$, we can write

\begin{equation}
\overline {\delta v}  = \frac{c}{{\lambda _0 }}\left( {\overline {\Delta \lambda }  - \overline {\Delta \lambda _{0ff} } \,\left( {1 + \frac{{\overline {v^{Ref} } }}{c}} \right)} \right).
\label{equ:1}
\end{equation}

We measure $\overline{\Delta \lambda}$ by taking the average of the distribution
of wavelength differences between the two lines .
Then, using equation (\ref{equ:1}) we calculate the relative velocity
$\overline {\delta v}$ of hot coronal ions with respect to Fe~{\sc x}~$\lambda$184
that we assume to have an average Doppler velocity $\overline{v^{Ref}}$
equal to the average Doppler velocity of Mg~{\sc x}~$\lambda$625
($-1.8 \pm 0.6~\rm km~s^{-1}$).

\begin{figure}[htp]
   \centering
   \includegraphics[width= 0.40 \paperwidth]{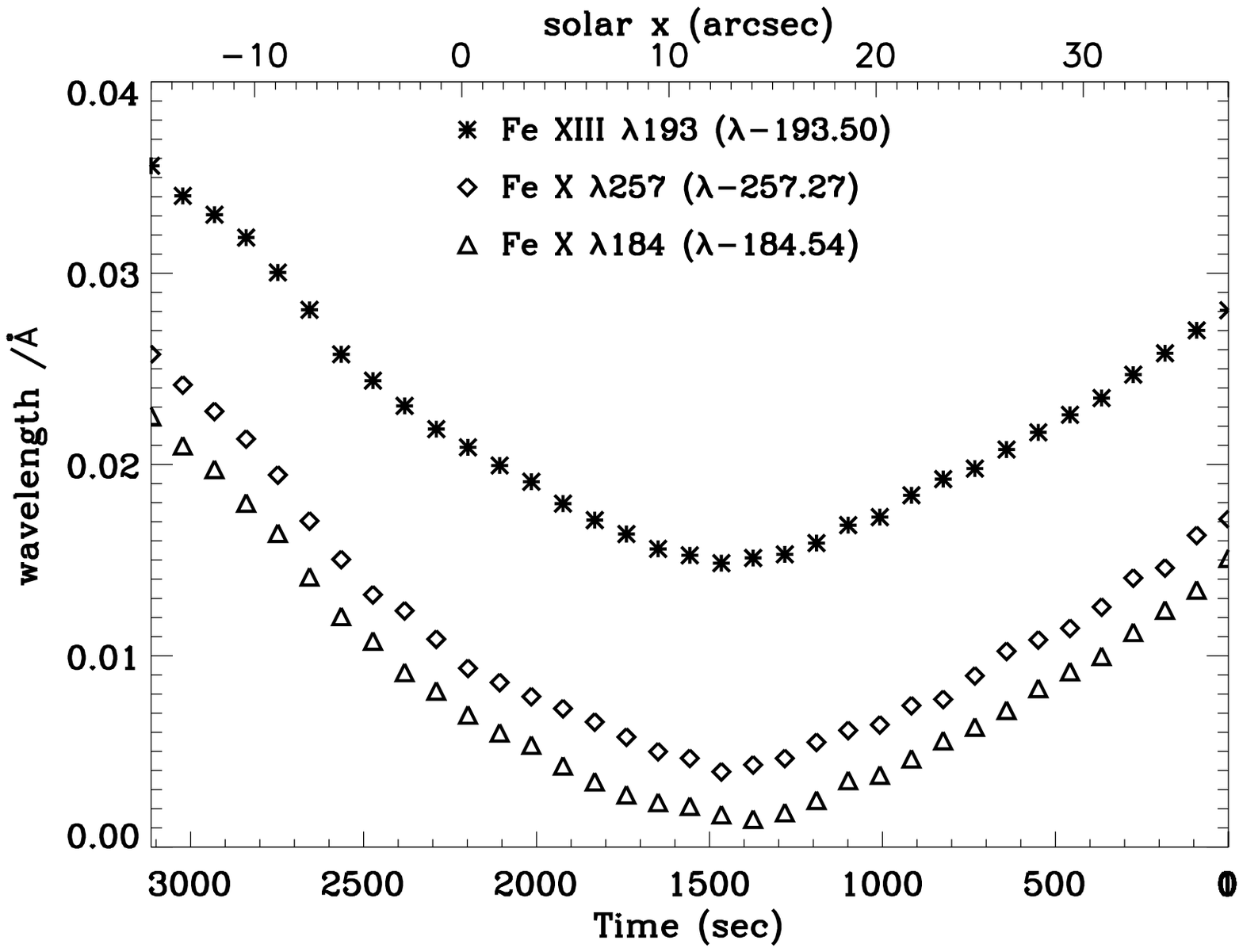}
   \includegraphics[width= 0.40 \paperwidth]{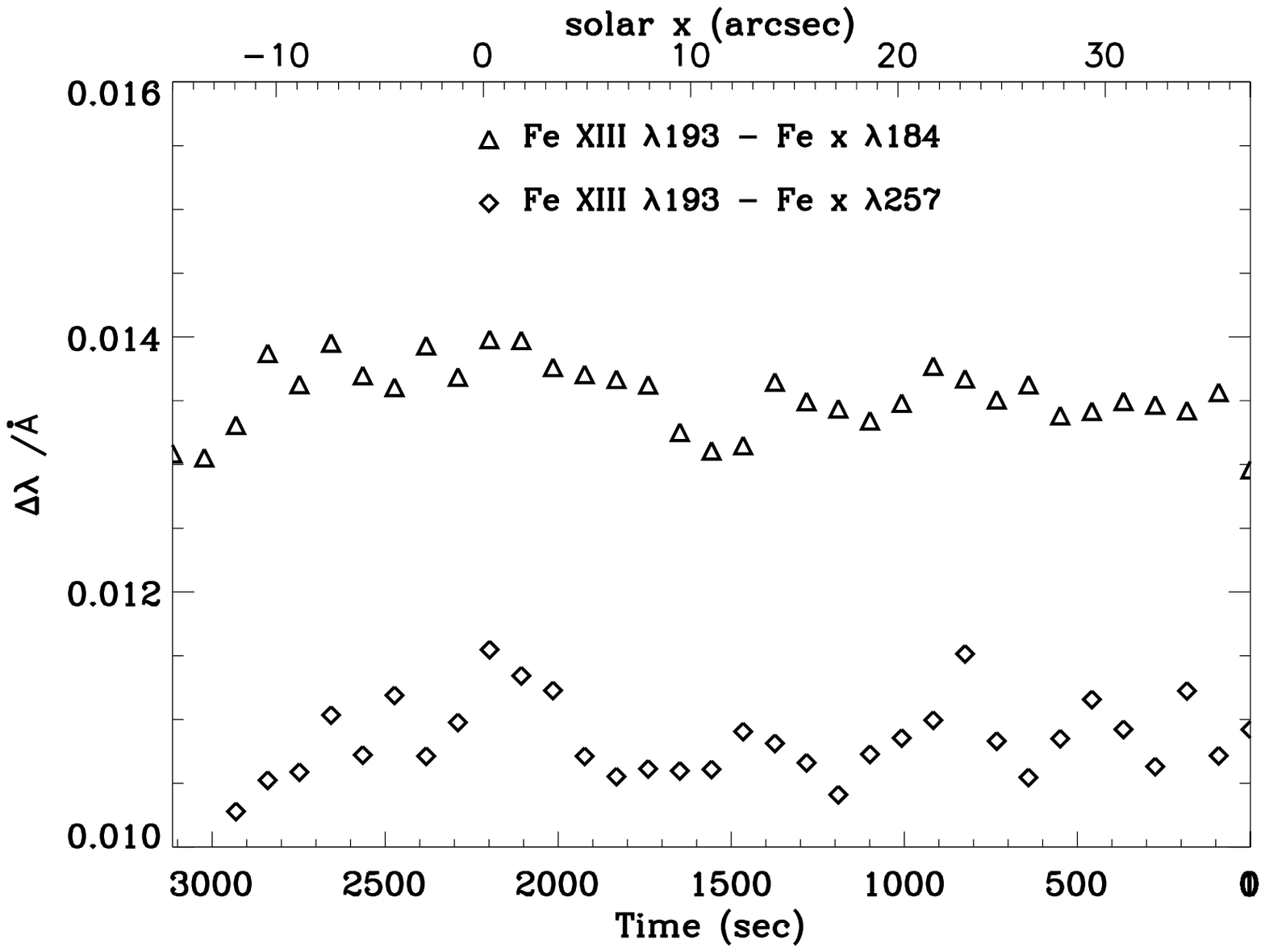}
     \caption{Upper panel: line positions, in wavelength units, of the
              Fe~{\sc xii}~$\lambda$193,
              Fe~{\sc x}~$\lambda$257, and Fe~{\sc x}~$\lambda$184
              lines versus time (and solar X) at disk centre.
            The wavelength values indicated in the panel
              have been subtracted to ease comparison.
              Lower panel: wavelength difference between spectral lines for
              the data in the upper panel versus time (and solar X).
              It can be seen that changes in line distances (wavelength
              differences) are very small ($\approx$1 m\AA,
              $\approx$ 1.5~km~s$^{-1}$) and show no dependence
              on time.}
   \label{16dec}
 \end{figure}

Because of thermal variation along the 98.5~min Hinode orbit, there is a shift
in the spectral line positions with time that needs to be corrected before using
EIS data. Routines for correcting this variation are provided with the data
reduction package and the correction has been applied to our data.
However, we are not sure that the above correction (based on the data itself)
is reliable down to the accuracy needed for the measurements discussed here
($\approx$ 2~km~s$^{-1}$).
A major advantage of the technique presented in this paper is that we
do not rely on this correction because we are just measuring the spectral
distances between spectral lines \footnote{This method also
ensures that there is no bias from the observed line intensities in the derived
average velocity.}. We only need to make sure that all
spectral lines move across the detectors in the same way, i.e., that there
are no changes in the image scale with time.
The top panel of Fig.~\ref{16dec} shows the dependence on time (and solar X)
of line positions (averages along the slit, in wavelength units) for three lines
(Fe~{\sc x}~$\lambda$184, Fe~{\sc x}~$\lambda$257, and
Fe~{\sc xii}~$\lambda$193) when no correction for temporal drifts is introduced.
Offsets in wavelength (indicated in the figure) have been subtracted to make the
comparison more evident.
The behaviour of all the lines is very similar.
The bottom panel of the figure shows line distances (differences in the values
in the top panel) versus time and solar X. It can be seen that changes in line
distance (wavelength differences) are very small ($\approx$1~m\AA,
$\approx$ 1.5~km~s$^{-1}$) and show no significant
dependence with time. Hence our assumption is fulfilled.

To improve the signal-to-noise ratio, before fitting the line profiles
we have binned the EIS on-disk spectra over three raster positions and over
three pixels along the slit (3$\times$5 binning for the Fe~{\sc xv}
$\lambda284$, weak on the quiet Sun). A 2$\times$2 binning was used for
the off-limb spectra.
 \section{Results}
Using the technique discussed in Section~\ref{sec:method} we have obtained the line of sight
velocities from the Doppler shifts of many coronal lines formed at temperatures between 1~MK
and 2~MK in the quiet Sun.
These lines, their deduced relative (to Fe~{\sc x} $\lambda$184)
Doppler shifts and the corresponding uncertainties
are listed in Table \ref{table:2}. In this table, we report the values obtained in
the four EIS raster scans together with the estimated errors.
Errors are obtained by error propagation analysis of Eq.~\ref{equ:1}, where the error on
$\overline {\Delta \lambda _{0ff}}$ is given by the width $\sigma$ of the off-limb
line-distance distribution. On the other hand, the on-disk line distance distribution
is also broadened by the different Doppler speeds characterising the two lines at different
locations.
For this reason the error for $\overline {\Delta \lambda}$ is assumed to be equal to that of
$\overline {\Delta \lambda _{0ff}}$, leading to the same error for a given line in all the
rasters.
An independent estimate of these errors can be obtained by looking at the
standard deviation of the measured values of $\overline {\delta v}$ obtained in the
four rasters. From Table \ref{table:2}, values of 0.1~km~s$^{-1}$ and 0.7~km~s$^{-1}$
would be obtained for Fe~{\sc xii}~$\lambda$195 and Fe~{\sc xv}~$\lambda$284,
respectively. These values are lower than our estimated errors and indicate the
most likely we do not underestimate the errors provided in
Table~\ref{table:2}.

\begin{figure*}
  \resizebox{17cm}{!}{\includegraphics{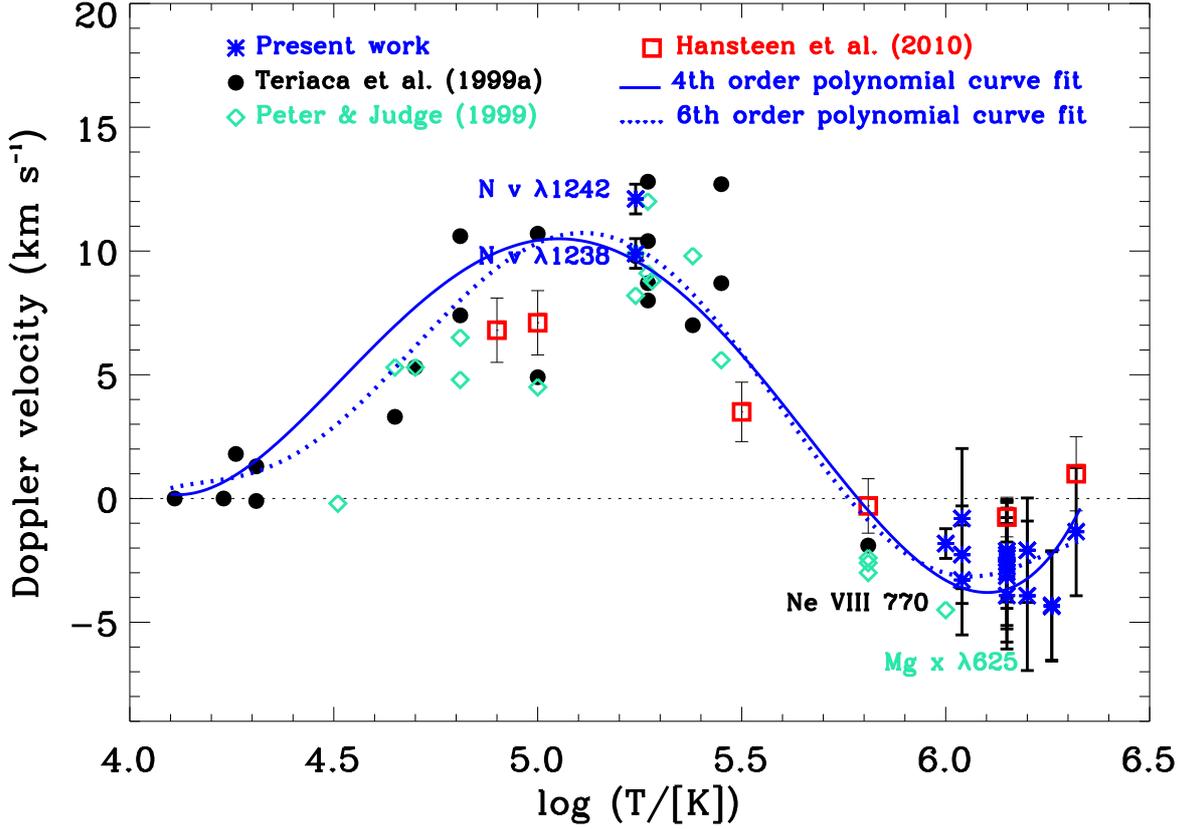}}
\caption{Average Doppler shift in the quiet Sun at disk centre of various TR
            and coronal ions measured from SUMER and EIS spectra.
          Positive values indicate red shifts (downflows) while negative
          values indicate blue shifts (upflows).
            Solid and dashed lines respectively represent fourth and
          and sixth order polynomial curves by-eye fitted to the combination of the
          measurements obtained in this work and
            those of \citet{Teriaca99a}.}
  \label{pic5}
 \end{figure*}

\begin{figure}
  \hspace{-0.3cm}\resizebox{8.5cm}{!}{\includegraphics{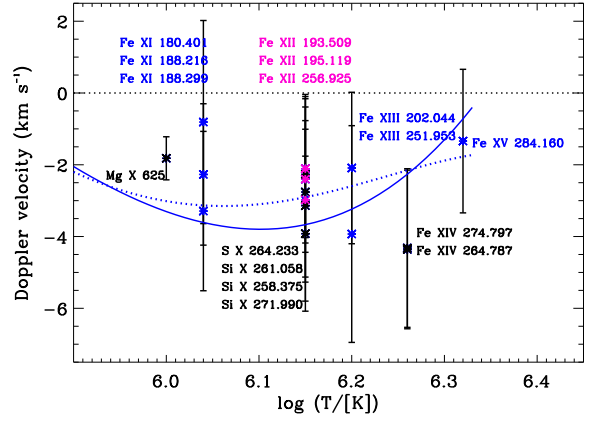}}
\caption{Magnification of the high-temperature part of Fig.~\ref{pic5} showing
the results obtained in this work for coronal lines.
         Symbols and lines as in Fig.~\ref{pic5}.}
  \label{pic6}
 \end{figure}

In the following, we take the average of rasters \#~3 and \#~4 as
representative values of the average Doppler shift on the quiet Sun, the
closest in time to the SUMER observations. By adding the value of average
Doppler velocity of our reference line (Mg~{\sc x}~$\lambda625$) to these
values ($\overline{v}=\overline{v^{Ref}}+\overline{\delta v}$), the absolute
Doppler velocity of lines are calculated (see Table~\ref{table:3}).
The error reported in this table, $\delta (\overline{v})$, is obtained by
summing in quadrature the errors on relative line Doppler shifts (reported in
Table~\ref{table:2}), and the error in measuring the average Doppler velocity
of our reference line (Mg~{\sc x}~$\lambda625$).
The values in Table~\ref{table:3}) are plotted in Fig. \ref{pic5}
and as a blow-up for temperatures above 1~MK in Fig. \ref{pic6}.

Figure \ref{pic5} displays the Doppler shift in the quiet
Sun at disk centre of various TR and coronal ions in
SUMER and EIS spectra. Positive values indicate red shifts (downflows), while
negative values indicate blue shifts (upflows). Values obtained in this
investigation (both SUMER and EIS) are marked by stars. In the figure the
results of \citet{Teriaca99a}, \citet{PetJud99}, and the theoretical
Doppler shifts of \citet{Hansteen2010} are also shown.

The solid and dashed lines respectively represent a by-eye fourth and sixth
order polynomial representation of the combination of our present work for
TR and coronal lines ($T\leq$ 2~MK) and those of \citet{Teriaca99a}.
Thus, the $v$~$versus$ log($T/$[K]) behaviour can be parametrised by the
coefficients that are listed in Table~\ref{table:4}.

At temperatures higher than 1~MK our results show a modest but clear
increase of the average blue shift with respect to the value measured around
1~MK ($-1.8$~km~s$^{-1}$), reaching a maximum of about $-4.4$ km~s$^{-1}$
at the Fe~{\sc xiv} formation temperature of 1.8 MK. At even higher
temperatures, the average line shift appears to decrease again, reaching a
value of $(-1.3\pm2.6)$¬km~s$^{-1}$ obtained in Fe~{\sc xv} $\lambda$284
line at 2.1~MK.

To check if our results are dependent on the choice of our reference line
(Fe~{\sc x}~$\lambda$184), we have used another Fe~{\sc x} line
($\lambda$257) and calculated all
Doppler velocities with respect to it. The Doppler velocities are found to
agree within $1~\rm km~s^{-1}$ with those in Table \ref{table:2}.

\begin{figure}
  \hspace{-0.3cm}\resizebox{8.5cm}{!}{\includegraphics{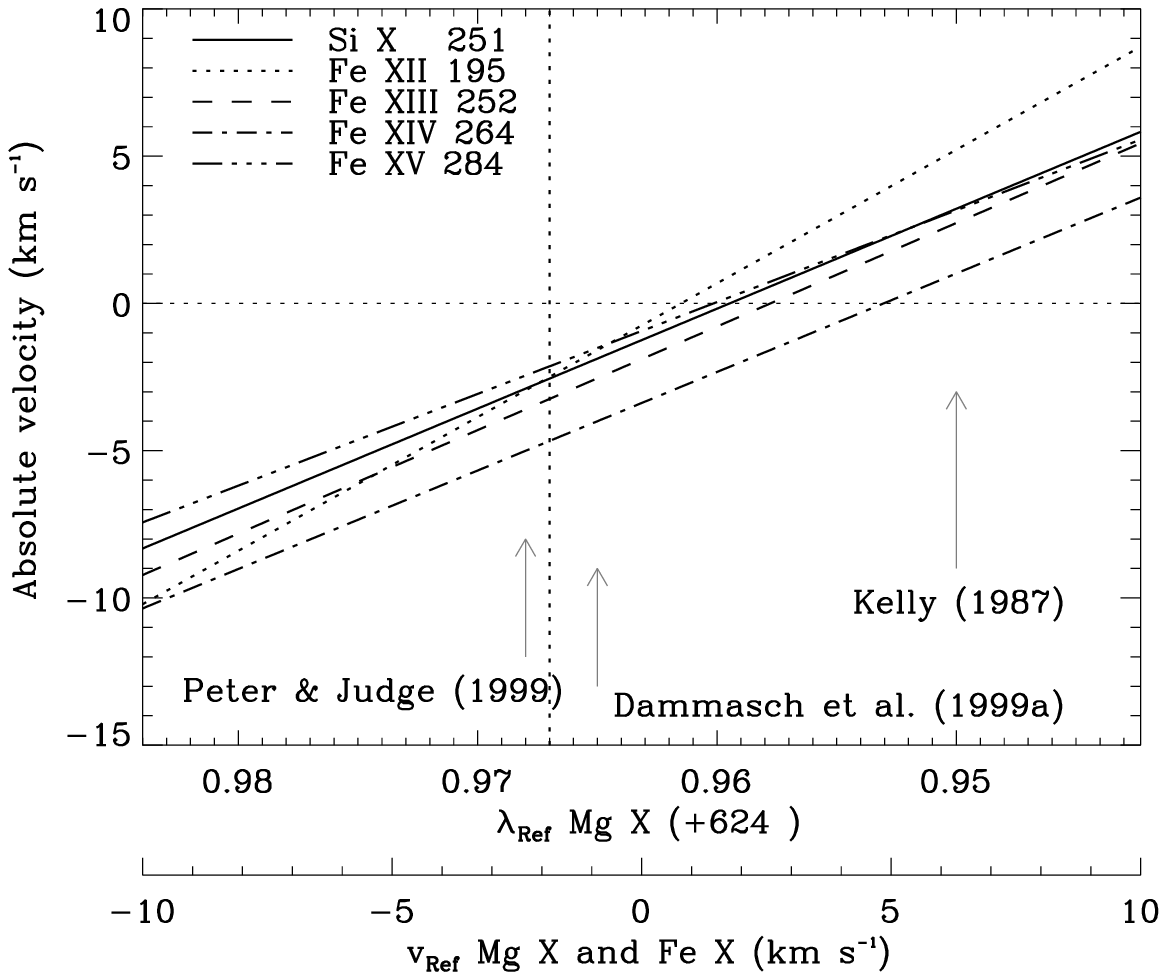}}
  \caption{ Variation in the absolute Doppler velocity of hot coronal ions
               as a function of the rest wavelength of the Mg~{\sc x} line and,
             consequently, of the Doppler velocity from
               the Fe~{\sc x}~$\lambda$184 reference line. The vertical dotted
             line corresponds to the adopted rest wavelength of 624.967~\AA.}
  \label{pic7}
 \end{figure}

It is clear that the absolute values of the velocities derived here depend
on the value adopted for the reference line (in this case $-$1.8~km~s$^{-1}$).
To check the dependence of our results on the value of the velocity of our
reference line (Fe~{\sc x}~$\lambda$184, which we have supposed to have the
same average Doppler velocity of Mg~{\sc x}~$\lambda$625), we used
Eq.~\ref{equ:1} to calculate the absolute velocity for different values of the
rest wavelength of the Mg~{\sc x} line (which ultimately determine the
velocity in the reference line).
The dependence of the Doppler shift on the Mg~{\sc x}
rest wavelength and on the resulting Doppler velocity is shown in
Fig.~\ref{pic7} for the Si~{\sc x}~$\lambda$261, Fe~{\sc xii}~$\lambda$195,
Fe~{\sc xiii}~$\lambda$251, Fe~{\sc xiv}~$\lambda$264, and
Fe~{\sc xv}~$\lambda$284 lines.
It should be noted that using the rest wavelength from \citet{Kelly87},
would result in red shifts for most of the coronal lines studied here.

\begin{table}
\caption{Doppler velocity of ions with respect to Fe~{\sc x}~$\lambda$184.}
\label{table:2}
\centering
\begin{tabular*}{8.8cm}{c c c c c c}
\hline \hline & \mbox{}\\[-1.5ex]

line                & \multicolumn{4}{c}{$\overline {\delta v}$} &$\delta(\overline {\delta v})$\\
                    & \multicolumn{4}{c}{[km~s$^{-1}$]} & [km~s$^{-1}$]  \\
                    &  \#~1 & \#~2 & \#~3 &  \#~4 & \\
\hline & \mbox{}\\[-1.5ex]
    Fe~{\sc x}~257.262   &    0.86 & 0.28 & 0.52 & 1.44 & 2.30   \\[1.5ex]
    Fe~{\sc xi}~180.401  &    0.96 & 1.14 & 0.76 & 1.22 & 2.77   \\
    Fe~{\sc xi}~188.216  &    0.15 &-0.71 &-0.25 &-0.68 & 1.88   \\
    Fe~{\sc xi}~188.299  &   -1.35 &-2.35 &-1.41 &-1.57 & 2.14   \\[1.5ex]
    Si~{\sc x}~258.375   &   -0.41 &-0.87 &-1.57 &-1.10 & 2.04   \\
    Si~{\sc x}~261.058   &   -0.96 &-0.04 &-1.19 &-0.73 & 2.29   \\
    Si~{\sc x}~271.990   &   -1.51 &-1.62 &-2.06 &-2.17 & 2.07   \\
    ~S~{\sc x}~264.233   &    0.21 &-0.01 &-0.47 &-0.47 & 2.09   \\[1.5ex]
    Fe~{\sc xii}~193.509  &  -0.64 &-0.83 &-0.16 &-0.46 & 1.98 \\
    Fe~{\sc xii}~195.119  &  -0.52 &-0.67 &-0.52 &-0.67 & 1.52 \\
    Fe~{\sc xii}~256.925  &  -0.54 &-1.00 &-1.70 &-0.65 & 2.76   \\[1.5ex]
    Fe~{\sc xiii}~202.044  & -0.43 &-0.73 &-0.14 &-0.43 & 2.02\\
    Fe~{\sc xiii}~251.953  & -1.54 &-3.08 &-2.84 &-1.42 & 2.96 \\[1.5ex]
    Fe~{\sc xiv}~264.787   &  -2.62 &-2.96 &-2.27 &-2.84 & 2.13\\
    Fe~{\sc xiv}~274.203   &  -2.47 &-2.14 &-3.45 &-1.59 & 2.13\\[1.5ex]

    Fe~{\sc xv}~284.160    &   1.38 & 0.33 & 1.28 &-0.31 & 2.56\\[1.5ex]
\hline
\end{tabular*}
\begin{tablenotes}
\item  Columns 2 to 5 list the results for the four different rasters. The
last column provides the error, which is equal for all rasters (see text).
\end{tablenotes}
\end{table}

\begin{table}
\caption{Absolute average Doppler velocity of ions (averaged over rasters \#3 and \#4).}
\label{table:3}
\centering
\begin{tabular*}{8cm}{c c c c c}
\hline \hline & \mbox{}\\[-1.5ex]
line (log($T$/[K])) & $\overline {v}$~[km~s$^{-1}$] & $\delta (\overline {v})$~[km~s$^{-1}$]~$^{\rm a}$ \\
\hline & \mbox{}\\[-1.5ex]
    Fe~{\sc x}~184.536 (6.00)  &  -1.8~${^{\rm b}}$ & 0.6~${^{\rm b}}$ \\
    Fe~{\sc x}~257.262 (6.00)  &  -0.82 & 2.38\\[1.5ex]
    Fe~{\sc xi}~180.401 (6.04) &  -0.81 & 2.83\\      
    Fe~{\sc xi}~188.216 (6.04) &  -2.27 & 1.97\\
    Fe~{\sc xi}~188.299 (6.04) &  -3.29 & 2.22\\[1.5ex]
    Si~{\sc x}~258.375 (6.15)  &  -3.14 & 2.13  \\
    Si~{\sc x}~261.058 (6.15)  &  -2.76 & 2.37  \\
    Si~{\sc x}~271.990 (6.15)  &  -3.92 & 2.16  \\
    ~S~{\sc x}~264.233 (6.15)  &  -2.27 & 2.17  \\[1.5ex]
    Fe~{\sc xii}~193.509 (6.15) & -2.11 & 2.07   \\
    Fe~{\sc xii}~195.119 (6.15) & -2.40 & 1.63  \\
    Fe~{\sc xii}~256.925 (6.15) & -2.98 & 2.82\\[1.5ex]
    Fe~{\sc xiii}~202.044 (6.20)& -2.09 & 2.11  \\
    Fe~{\sc xiii}~251.953 (6.20)& -3.93 & 3.02 \\[1.5ex]
    Fe~{\sc xiv}~264.787 (6.26) & -4.36 & 2.21  \\
    Fe~{\sc xiv}~274.203 (6.26) & -4.32 & 2.21  \\[1.5ex]

    Fe~{\sc xv}~284.160 (6.32)  & -1.34 & 2.63  \\[1.5ex]
\hline
\end{tabular*}
\begin{tablenotes}
\item $^{\rm a}$ The last column provides the error on the absolute average Doppler velocities (see text).
\item $^{\rm b}$ Absolute velocity and error assumed equal to that of Mg~{\sc x}.
\end{tablenotes}
\end{table}
\begin{table}
\caption{Coefficients of the fourth and sixth order polynomial representations
of the average Doppler velocity versus temperature behaviour of the quiet Sun.}
\label{table:4}
\centering
 \begin{tabular*}{6cm}{ccc}  
  \hline  \hline  & \mbox{}\\[-1.5ex]
coefficient & $4^{th}$ order &  $6^{th}$ order       \\
  \hline & \mbox{}\\[-1.5ex]
a$_{0}$ &  7874.8  & -185202.25 \\
a$_{1}$ & -6427.5  &  223219.47 \\
a$_{2}$ &  1943.26 & -111257.36 \\
a$_{3}$ & -257.766 &  29345.005 \\
a$_{4}$ &  12.6596 & -4319.1993 \\
a$_{5}$ &          &  336.35535 \\
a$_{6}$ &          & -10.827987 \\
   \hline
 \end{tabular*}
\begin{tablenotes}
\item[1] All decimals shown here are significant.
\end{tablenotes}
 \end{table}
%
 \section{Summary and conclusions}
In the present study we measured the Doppler shift of TR
and coronal lines on the quiet Sun (see Tables 1 \& 3).
We measured the absolute average velocity of Mg~{\sc x}~$\lambda$625,
N~{\sc v}~$\lambda$1238, and N~{\sc v}~$\lambda$1242 lines using SUMER data
obtaining a blue shifted Doppler velocity of ($-1.8 \pm 0.6$)~km~s$^{-1}$ for
Mg~{\sc x}~$\lambda$625 and red shifts (downward motions) of ($9.9 \pm 0.6$) and
($12.1 \pm 0.6$) for the N~{\sc v}~$\lambda$1238 and $\lambda$1242 lines.
Our results for TR lines agree with previous
studies done with SUMER \citep{Brekke97, Chae98, PetJud99, Teriaca99a}.
Our result for the Mg~{\sc x} line is also consistent with the
literature, once the differences on the adopted rest wavelength are accounted
for (see Section 2.1). Of course, our value for the average shift in the
Mg~{\sc x} line depends on the adopted rest wavelength of 624.967~\AA~
\citep[average between the values of]{Dam99a, PetJud99}. Since the
velocities obtained from the EIS lines also depend on the absolute velocity of
the Mg~{\sc x} ions, the precise determination of the rest wavelength of
Mg~{\sc x} is paramount. The accurate, independent, and highly
consistent measurements of \citet{Dam99a} and \citet{PetJud99} are the best
available so far. In any case, Fig.~\ref{pic7} allows assessing the effect of
a different rest wavelength for the Mg~{\sc x} $\lambda625$ line
on the resulting Doppler shifts.

By using cospatial and nearly cotemporal SUMER \& EIS data we,
for the first time, measured the average Doppler shift
of several coronal lines with formation temperatures above 1~MK in the quiet Sun near
disk centre. We obtained negative Doppler shifts (upward motions) between
-2 to -4 km~s$^{-1}$ at temperatures between 1 and 1.8~MK.
At the highest temperature we can investigate, $T=2.1$¬MK (Fe~{\sc xv} $\lambda284$),
we find indications of a reduction of the observed blue shift down to 1.3 km~s$^{-1}$,
a value that is compatible with no line shift, but also with a blue shift of up to 4 km~s$^{-1}$.

As discussed in the Introduction, 1-D models assuming energy release in
magnetic loops had successfully managed to reproduce the observer red shift
at TR temperatures and blue shifts at coronal temperatures
\citep[e.g.,][]{Teriaca99b, Spadaro-etal:06}.
However, these models do not explicitly consider the heating mechanism.

More recently, comprehensive 3-D models of the solar
atmosphere have been developed. Models assuming that coronal heating is
caused by Joule dissipation of currents produced by stressing and braiding of the magnetic field
produce red shifts at all temperatures \citep{Peter2004, Peter2006}, which is
contrary to the present results and those of \citep{PetJud99} and \citep{Teriaca99a}.

The addition to the model of episodic injections of emerging magnetic flux
that reconnects with the existing field, produces
rapid, episodic heating of the upper chromospheric plasma
to coronal temperatures \citep{Hansteen2010}.
Doppler shifts are computed by these authors for a model with an unsigned
magnetic field of $\mid B_{z}\mid \approx$ 75~G, which is compatible with
the quiet Sun
\citep{2004Natur.430..326T, 2006ApJ...636..496D, 2010A&A...513A...1D}.
The resulting Doppler shifts are roughly consistent with those reported
here at all temperatures up to the Fe~{\sc xiv} formation temperature
(log($T$/[K])=6.26) (see Fig.~\ref{pic5}).
Above this value, the line shifts predicted by the model return to
zero and become
slightly red shifted at the Fe~{\sc xv} formation temperature
(log($T$/[K])=6.32).

Our Fe~{\sc xv} observations, instead, yield a blue shifted value.
However, the error bar is large enough that no firm conclusion can be
drawn.
A real discrepancy at this temperature between the observations and the model
may possibly be due to boundary effects in the adopted computing box or, more
simply, to the fact that the turning point to zero or red shifted values occurs
in reality at temperatures even higher than those measurable in this study.
However, it may also be possible that important
ingredients are still missing in the theoretical/numerical description.

\citet{Hansteen2010} suggest that the high speed flows generated by the
localised heating events may be related to the high speed upflows that have
been deduced from significant blueward asymmetries in Hinode/EIS observations
of many TR and coronal lines \citep{Hara-etal:2008, DePontieu-etal:2009,
McIntosh-etal:2009a, McIntosh-etal:2009b} that have been linked to spicules
\citep[]{DePontieu-etal:2011}. Thus, a more accurate introduction of the
effect of spicules into the model may be necessary.

The model also foresees a smaller amount of Doppler shifts at both TR (lower
red shifts) and coronal temperatures (lower blue shifts).
However, this is not surprising (as the model does not specifically
attempt to reproduce the solar region studied here), and the model is most
likely compatible with our measurement.
 Since the model does not consider open field lines, this would indicate
that the observed line shifts are simply a by-product
of the heating of, and related mass supply to, closed loops.
This would be consistent with the finding that the plasma
outflows seen on the quiet Sun, such as the Ne~{\sc viii} blue shift, are largely
associated with closed loops \citep{2008A&A...478..915T}.

 However, it cannot also be excluded that hot open funnels
significantly contribute to the blue shifted emission observed between 1
and 2 MK.
Only a detailed analysis of spatially resolved flows compared to
field extrapolations \citep[similar to what was done for the Ne~{\sc viii} line at lower
temperatures by][]{2008A&A...478..915T} would shed some light on the question.
Given that these hotter lines are weak on the quiet Sun,
this analysis is very difficult and will require a VUV spectrometer
with much higher throughput than currently available, such as the spectrometer
proposed for the Japanese Solar C mission \citep{Teriaca-etal:2011}.

\begin{acknowledgements}
We thank Giulio Del Zanna, Suguru Kamio, and Maria Madjarska
for useful discussions and comments. We also thank the SUMER and
Hinode Teams for their support in obtaining the data.
We finally thank the
anonymous referee for useful comments and suggestions.
This work has been partially supported by WCU grant
No. R31-10016 funded by the Korean
Ministry of Education, Science and Technology.
The SUMER project is financially supported by DLR, CNES, NASA, and the ESA
PRODEX programme (Swiss contribution). SOHO is a mission of international
cooperation between ESA and NASA.
Hinode is a Japanese mission developed and launched by ISAS/JAXA,
collaborating with NAOJ as a domestic partner, and NASA and STFC (UK) as
international partners. Scientific operation of the Hinode mission is
conducted by the Hinode science team organised at ISAS/JAXA. This team
mainly consists of scientists from institutes in the partner countries.
Support for the post-launch operation is provided by JAXA and NAOJ (Japan),
STFC (U.K.), NASA, ESA, and NSC (Norway).
Neda Dadashi acknowledges a PhD fellowship of the International Max Planck
Research School on Physical Processes in the Solar System and Beyond.
\end{acknowledgements}

\bibliographystyle{aa}

\bibliography{dadashi}

\end{document}